\documentclass[11pt]{article}
\usepackage{latexsym}  
\usepackage{amssymb}
\usepackage{graphicx}
\usepackage{amsmath}

\begin{document}


\hspace{10cm}{OU-HET-687/2010}

\vspace{3mm}

\begin{center}
{\Large\bf Yukawaon Model with U(3)$\times$O(3) Family Symmetries}

\vspace{3mm}
{\bf Yoshio Koide}

{\it Department of Physics, Osaka University,  
Toyonaka, Osaka 560-0043, Japan} \\
{\it E-mail address: koide@het.phys.sci.osaka-u.ac.jp}

\date{\today}
\end{center}

\vspace{3mm}
\begin{abstract}
A quark and lepton mass matrix model with family symmetries 
U(3)$\times$O(3) is investigated on the basis of the 
so-called yukawaon model. 
In the present model, quarks and leptons are assigned
to $(\ell, e^c, u^c) \sim ({\bf 3}, {\bf 3}, {\bf 3}^*)$ of U(3) and 
$(q, d^c, \nu^c) \sim ({\bf 3}, {\bf 3}, {\bf 3})$ of O(3).  
Then, the neutrino mass matrix is given by 
$M_\nu = m_D M_R^{-1} m_D^T$ with $m_D \propto \langle \Phi_e \rangle$,
where the charged lepton mass matrix $M_e$ is given by
$M_e = k \langle \Phi_e \rangle \langle \Phi_e^T \rangle$.
A merit in considering U(3)$\times$O(3) lies in that
we can lower the cutoff scale $\Lambda$ in the yukawaon
model.
\end{abstract}

\vspace{5mm}

{\large\bf 1. Introduction}

One of the most challenging problems in contemporary 
particle physics is to clarify the origin of flavors.
For such a purpose, it is interesting to investigate
whether the observed flavor physics phenomena can 
be understood from a concept of a family gauge
symmetry or not.
The present data \cite{PDG10} suggest that
numbers of lepton- and quark-families are both three.
Then, from a point of view of a unification model
of quarks and leptons, it will be natural
to consider that the quarks and leptons obey the same
family symmetry.
However, at present, this is experimentally not
yet confirmed. 
Can quarks and leptons be described by a sole 
family symmetry? 
In this paper, we investigate a possibility that
quarks and leptons obey different family symmetries
from each other.

In the present paper, by assuming family symmetries 
U(3)$\times$O(3), we will propose a new version 
of the so-called ``supersymmetric yukawaon" model 
\cite{yukawaon,O3_PLB09} (a kind of ``flavon" model
\cite{flavon}).
In the yukawaon model, all effective 
Yukawa coupling constants $Y_f^{eff}$  ($f=u,d,e,\cdots$) 
are given by vacuum expectation values (VEVs) of 
``yukawaons" $Y_f$ as 
$$
Y_f^{eff} = y_f \frac{\langle Y_f \rangle}{\Lambda} .
\eqno(1.1)
$$
The yukawaons $Y_f$ are singlets under the conventional
gauge symmetries SU(3)$_c \times$SU(2)$_L\times$U(1)$_Y$,
and have only family indices.
For $\mu < \Lambda$, the effective Yukawa coupling constants
$Y_f^{eff}$ evolve as those in the standard model. 
The effective Lagrangian is practically identical with
the minimal SUSY standard model (MSSM) \cite{MSSM} except 
for that $Y_f$ are not constants, but superfields.  
(A brief review of the yukawaon model is given in Sec.2.)

In the present model with family symmetries 
U(3)$\times$O(3), we assume 
the following would-be Yukawa interaction terms:
$$
W_Y = \frac{y_e}{\Lambda} {\ell}_i Y_e^{ij} e^c_j H_d 
+ \frac{y_\nu}{\Lambda} {\ell}_i \Phi_e^{i \alpha}
\nu^c_\alpha H_u 
+\lambda_R \nu^c_\alpha Y_R^{\alpha\beta} \nu^c_\beta 
+ \frac{y_u}{\Lambda} u^{c i} Y^u_{i\beta} q_\beta H_u 
+ \frac{y_d}{\Lambda} d^c_\alpha Y^d_{\alpha\beta} 
q_\beta H_d ,
\eqno(1.2)
$$
together with the conventional Higgs field term 
$W_H = \mu_H H_u H_d$.
Here, $\ell$ and $q$ are defined by $\ell=(\nu_L, e_L)$ and 
$q=(u_L, d_L)$, and indexes $i,j,\cdots$ and 
$\alpha, \beta, \cdots$ denote those of U(3) and O(3), respectively.
Note that, as seen in Eq.(1.2), the yukawaon $\Phi_e$ 
plays a role of the substitute for a yukawaon $Y_\nu$.
This is the most characteristic in the present model.
The neutrino Dirac mass matrix $m_D$ is given by 
$$
m_D = \frac{y_\nu}{\Lambda} \langle \Phi_e\rangle 
\langle H_u^0\rangle .
\eqno(1.3)
$$
As we show later, fields $Y_e$, $Y_u$, $\cdots$ satisfy the  
following VEV relations:
$$
\langle Y_e^{ij} \rangle \propto 
\langle \Phi_e^{i\alpha} \rangle
\langle \Phi_e^{T\alpha j} \rangle , \ \ \ 
\langle Y_u^{i\alpha} \rangle \propto 
\langle \Phi^u_{i\beta} \rangle
\langle \bar{E}^{T\beta j}\rangle
\langle \Phi^u_{j \alpha} \rangle ,
\eqno(1.4)
$$
where $\langle \bar{E}^{i \alpha} \rangle = v_E \delta^{i\alpha}$.

So far, in a series of yukawaon models, the flavor symmetry 
was either U(3) \cite{U3_yukawaon} or O(3) \cite{O3_PLB08,O3_PLB09}, 
and it was a global symmetry. 
In general, when the family symmetry is global, 
unwelcome massless scalars appear in the model.
Therefore, in this paper, we want to consider that the family 
symmetry is local.
However, since there are many family symmetry non-singlet 
fields in the yukawaon model, the family gauge symmetry cannot 
be asymptotic free, so that it is feared that the gauge coupling 
constant bursts at $\mu = \Lambda$. 
(Therefore, so far, we have not consider a possibility that 
the family symmetry in the yukawaon model is local.)
If we consider a model with different family symmetries 
for quarks and leptons, we will be able to soften such a 
trouble. 

The present model with U(3)$\times$O(3) symmetries
has received a hint from a charged lepton mass matrix model 
with U(3)$\times$O(3) symmetries which has recently been 
proposed by Sumino \cite{Sumino09PLB,Sumino09JHEP}.  
In the Sumino model, the charged lepton mass term is 
generated by a would-be Yukawa interaction 
$$
H_e = \frac{y_e}{\Lambda^2} 
\bar{\ell}_{L}^i \Phi^e_{i \alpha}
 \Phi^{eT}_{\alpha j} e_{R}^j H ,
\eqno(1.5)
$$
where $H$ is the Higgs scalar in the standard non-SUSY model.
(Sumino' model has not been based on a SUSY scenario.)
The charged lepton masses $m_{e_i}$ are acquired from  
the vacuum expectation value (VEV) of the scalar $\Phi^e$
\cite{Koide90MPL},
i.e. they are given by $m_{ei}=(y_e/\Lambda^2) 
\langle (\Phi^e)_{i\alpha} \rangle \langle (\Phi^{eT})_{\alpha i} 
\rangle \langle H^0 \rangle$.
In other words, the VEV of $\Phi^e$ has a form 
$\langle \Phi^e \rangle_e \propto  {\rm diag}(\sqrt{m_e}, 
\sqrt{m_\mu}, \sqrt{m_\tau})$, where the suffix ``$e$" 
denotes that a VEV matrix $\langle A\rangle$ takes
a form $\langle A\rangle_e$ in  a flavor basis 
in which the charged lepton mass matrix $M_e$ is diagonal. 
However, in his model, O(3) is not a family
symmetry.
Besides, Sumino has mentioned nothing about quark and 
neutrino family assignments explicitly.
In this paper, we regard the O(3) symmetry as 
another family symmetry which is related to quarks and
neutrinos.

In the Sumino model, it is essential that the left- and right-handed
charged leptons $e_{Li}$ and $e_{Ri}$ are assigned to ${\bf 3}$
and ${\bf 3}^*$ of U(3) family symmetry, respectively. 
(A similar fermion assignment has been proposed by 
Applequist, Bai and Piai \cite{Appelquist06}.)
The reason for this assignment is as follow:
Sumino's interest is in the charged 
lepton mass relation~\cite{Koidemass}
$$
K \equiv \frac{m_e +m_\mu + m_\tau}{
   (\sqrt{m_e} + \sqrt{m_\mu} + \sqrt{m_\tau})^2} 
   = \frac{2}{3} . 
\eqno(1.6)
$$
The relation $K=2/3$ is satisfied with the order of $10^{-5}$ 
for the pole masses, i.e. 
$K^{pole}=(2/3)\times (0.999989 \pm 0.000014)$ 
\cite{PDG10}, while it is only valid with 
the order of $10^{-3}$ for the running masses, 
e.g. $K(\mu)=(2/3)\times (1.00189 \pm 0.00002)$ 
at $\mu =m_Z$. 
However, in conventional mass matrix models, ``mass" 
means not  ``pole mass" but ``running mass." 
Sumino has seriously taken why the mass formula 
$K=2/3$ is so remarkably satisfied with the pole masses.  
The deviation of $K(\mu)$ from $K^{pole}$ is 
caused by a logarithmic term $m_{ei}\log(\mu/m_{ei})$ 
in the radiative correction term \cite{Arason} due to photon
$$
m_{ei}(\mu) = m_{ei}^{pole}\left[ 1-\frac{\alpha(\mu)}{\pi}
\left(1 +\frac{3}{4} \log \frac{\mu}{m_{ei}(\mu)} \right)
\right].
\eqno(1.7)
$$
Therefore, he assumed that a family symmetry is local, 
and that the logarithmic term in the 
radiative correction due to photon is  canceled by 
that due to  family gauge bosons.
As a result, we can obtain $K(\mu) =K^{pole}$. 
(However, it does not mean $m_{ei}(\mu)=m_{ei}^{pole}$.
The cancellation takes place only for the term with 
$\log m_{ei}$). 

In this paper, stimulated by Sumino's idea, we consider 
a model with family symmetries U(3)$\times$O(3) and with 
the assignments $\ell \sim {\bf 3}$ and $e^c \sim {\bf 3}$
of U(3). 
However, the purpose of the present paper is not to give 
the observed charged lepton mass spectra, but to describe a unified
description of the quark and neutrino mass spectra and mixings by
using the observed charged lepton mass spectrum as the input 
parameters. 
For convenience, we will use the charged lepton mass values 
at the $\mu=m_Z$ as the input values, because it is not essential
for numerical predictions in quark and neutrino sectors. 

In this paper, according to Sumino's suggestion, we assume that O(3) 
is already completely broken at an energy scale $\mu=\Lambda$,  
so that all the O(3) gauge bosons become massive and decouple from 
the present effective theory below $\Lambda$. 

In the conventional yukawaon model, the neutrino Dirac mass 
matrix $m_D$ was ad hoc given by $m_D \propto M_e$ 
(i.e. $\langle Y_\nu \rangle \propto \langle Y_e\rangle$)
from a phenomenological point of view (see Sec.2).
Therefore, we are obliged to accept a cutoff energy scale 
$\Lambda \sim 10^{12}$ GeV from the neutrino phenomenology
(see Sec.5), 
so that most of new physics as to the yukawaons become 
invisible. 
In the present model, the yukawaon $Y_e$ in the charged 
lepton sector is given by $Y_e^{ij}$ [${\bf 6}^*$ of U(3)], 
while a yukawaon in the neutrino (Dirac) sector is given by 
$Y_\nu^{i\alpha}$
[$({\bf 3}^*,{\bf 3})$ of U(3)$\times$O(3)].
Therefore, we can regard the field $\Phi_e^{i\alpha}$ in 
the Sumino model as $Y_\nu^{i\alpha}$.
Then, as we state in Sec.5, in the present model with 
$m_D \propto \langle \Phi_e \rangle$, we can have a possibility that
$\Lambda$ takes a considerably low value (e.g. $\Lambda 
\sim 10^8$ GeV), and the lightest family gauge boson 
$A_1^1$ can have a mass of a few TeV.
This is the greatest merit in considering U(3)$\times$O(3)
family symmetries in the yukawaon model. 

As we see later, 
since we want that the phenomenological success in the 
previous yukawaon model \cite{O3_PLB09} is inherited in the 
present model, as far as numerical results are concerned, 
most of the numerical results in the present model will 
be the same as those in the old model and not new.
The differences of the present U(3)$\times$O(3) yukawaon model 
from the previous O(3) yukawaon model will be summarized in 
the end of the next section.





\vspace{2mm}

{\large\bf 2.  Brief review of the yukawaon model}

Although the yukawaon model is a kind of the flavon model 
\cite{flavon},
differently from the conventional flavon models, the quarks 
and leptons are assigned to ``triplets" (and/or ``anti-triplets")
of a non-Abelian group G, e.g. not to ${\bf 2}+{\bf 1}$ of
SU(2), ${\bf 1}+{\bf 1}' +{\bf 1}^{\prime\prime}$ of U(1)$^3$, 
and so on.
The VEV values of yukawaons $Y_f$ with $3\times 3$ 
($3\times 3^*$) of G are directly determined by a structure of
a scalar potential which is invariant under the symmetry G.

The yukawaon model intends to describe all quark and lepton 
mass matrices based on only a fundamental VEV matrix 
$\langle \Phi_e \rangle$.
In the supersymmetric yukawaon model, 
the VEV matrices $\langle Y_f\rangle$ are related to the 
fundamental VEV matrix $\langle \Phi_e \rangle$ by using
SUSY vacuum conditions.
We cannot always uniquely determine a superpotential form 
from a flavor symmetry alone. 
In the previous yukawaon model, in order to distinguish 
a yukawaon $Y_f$ from other yukawaons $Y_{f'}$,
we assigned ``sector" charges (U(1)$_X$ charges)  by hand. 
(For example, we assign the sector charges as $Q_X(Y_e)=x_e$, 
$Q_X(e^c)=-x_e$, $Q_X(Y_u)=x_u$, $Q_X(u^c)=-x_u$, and so on, 
in each sector $f=e,\nu,u,d$ and $Q_X(Y_R)=2 x_\nu$. 
We assign $Q_X=0$ to the SU(2) doublet fields.) 
In contrast to the previous model, in the present model, 
we do not need such a sector charge.
We can distinguish the yukawaons by U(3)$\times$O(3) 
assignments and $R$ charges. 
Besides, in order to forbid unwelcome terms with $\Lambda^{-n}$ 
($n\gg 1$), we need $R$ charge assingments.

The superpotential for yukawaons is usually given by a form
$$
W=\sum_A f_A (Y_f, Y_{f'}, \cdots) \Theta_A ,
\eqno(2.1)
$$
where $\Theta_A$ is an auxiliary superfield. 
Therefore, a SUSY vacuum condition $\partial W/\partial \Theta_A=0$
leads to a VEV relation $f_A (\langle Y_f \rangle,  
\langle Y_{f'} \rangle, \cdots)=0$.
We assume that our vacuum always takes 
$\langle \Theta_A \rangle =0$.
Then, other vacuum conditions $\partial W/\partial Y_f=0$ 
do not give any VEV relation, because each term in 
those equations always contains one $\Theta$ field.
For example, the VEV relation $\langle Y_e \rangle\propto 
\langle \Phi_e \rangle\langle \Phi_e^T \rangle$ in Eq.(1.4) is
derived from a superpotential
$$
W_e = (\mu_e Y_e^{ij} +\lambda_e \Phi_e^{i\alpha} \Phi_e^{T\alpha j}
)\Theta^e_{ji} .
\eqno(2.2)
$$
Therefore, a SUSY vacuum condition $\partial W/\partial \Theta^e =0$
($W=W_e+ \cdots$) leads to
$$
\langle Y_e \rangle = - \frac{\lambda_e}{\mu_e} 
\langle \Phi_e \rangle\langle \Phi_e^T \rangle .
\eqno(2.3)
$$
(The bilinear form (2.3) for the charged lepton mass matrix is 
needed for an explanation of the charged lepton mass relation
$K=2/3$ \cite{Koide90MPL}.)
Since we take a vacuum with $\langle \Theta^e \rangle =0$, 
other conditions $\partial W/\partial Y_e = \mu_e \Theta^e + \cdots =0$ 
and $\partial W/\partial \Phi_e = \lambda_e (\Phi_e^T \Theta^e +
\Theta^{eT} \Phi_e ) + \cdots =0$ do not affect the relation (2.3).
In the present model, although the model is supersymmetric, 
``SUSY" plays only a role in obtaining VEV relations among yukawaons.
What we practically investigate are only quarks and leptons 
as fermions and yukawaons as scalars.
Nevertheless, we cannot dispense with SUSY, because,
in a non-SUSY model, we cannot have such the convenient prescription 
with $\Theta$ fields given in Eq.(2.1). 
 
For the time being, we assume that the observed supersymmetry breaking 
is induced by a gauge mediation mechanism (not including family 
gauge symmetries), 
so that our VEV relations among yukawaons are still valid after
the SUSY was broken in the quark and lepton sectors.

In the previous yukawaon mode \cite{O3_PLB09}
(we refer to it as the O(3) model), 
the family symmetry O(3) was global. 
By using SUSY vacuum conditions, 
we could successfully obtain reasonable quark 
and lepton mass matrices, especially excellent 
predictions for up-quark mass ratios and neutrino 
mixing parameters by adjusting only two parameters.
In contrast to the O(3) model, 
in the present paper, since we assume U(3)$\times$O(3)
as family symmetries, the theoretical framework is
changed.
However, we want to inherit the phenomenological
success from the old yukawaon model.

In the O(3) model, a neutrino mass matrix $M_\nu$ is given by 
a seesaw-type mass matrix $M_\nu = m_D M_R^{-1} m_D^T$,
where the Dirac and Majorana mass matrices $m_D$ and 
$M_R$ are given by
$$
m_D \propto  M_e \propto \langle Y_e \rangle , 
\eqno(2.4)
$$
$$
M_R \propto \langle \Phi_u \rangle \langle P_u \rangle 
\langle Y_e \rangle +
\langle Y_e \rangle \langle P_u \rangle \langle \Phi_u \rangle
$$
$$
+\xi_\nu (\langle \Phi_u \rangle \langle Y_e \rangle 
\langle P_u \rangle 
+\langle P_u \rangle \langle Y_e \rangle \langle \Phi_u \rangle) 
+\xi_0 \Lambda \langle Y_e \rangle \langle Y_e \rangle ,
\eqno(2.5)
$$
respectively. 
Here, the last term ($\xi_0$ term) has been added in order to 
adjust the ratio $R_\nu \equiv \Delta m^2_{solar}/
\Delta m^2_{atm}$ without affecting neutrino mixing 
parameters, because $M_\nu \propto \langle Y_e\rangle [(\cdots ) 
+\xi_0 \langle Y_e\rangle \langle Y_e\rangle]^{-1} \langle Y_e\rangle
= [\langle Y_e\rangle^{-1}(\cdots)\langle Y_e\rangle^{-1} 
+ \xi_0 {\bf 1}]^{-1}$ and the mixing matrix for $M_\nu$ is 
identical to that for $M_\nu^{-1}$ except for the phase factors
\cite{inverse_Mnu}. 
(In the O(3) model, the yukawaon $Y_R$ has the same U(1)$_X$ 
charge as $Y_e Y_e$, i.e. $Q_X(Y_R)=2 Q_X(Y_e)$.)
The charged lepton mass matrix $M_e$ is given by 
$M_e \propto \langle Y_e \rangle = k_e \langle \Phi_e \rangle 
\langle \Phi_e \rangle^T$ ($k_e=-\lambda_e/\mu_e$),  
while the quark mass matrices $M_u$ and $M_d$ are given by 
$$
M_u^{1/2}\propto  \langle \Phi_u \rangle = 
k_u \langle \Phi_e \rangle ({\bf 1} + a_u X) 
\langle \Phi_e \rangle , \ \ \ 
M_d \propto \langle Y_d \rangle =
k_d \langle \Phi_e \rangle ({\bf 1} + a_d X) 
\langle \Phi_e \rangle ,
\eqno(2.6)
$$
where
$$
{\bf 1} =\left(
\begin{array}{ccc}
1 & 0 & 0 \\
0 & 1 & 0 \\
0 & 0 & 1
\end{array} \right) ,
\ \ \ 
X= \frac{1}{3} \left(
\begin{array}{ccc}
1 & 1 & 1 \\
1 & 1 & 1 \\
1 & 1 & 1
\end{array} \right) ,
\eqno(2.7)
$$
and $P_u$ is defined as a field with a VEV matrix form
$$
\langle P_u \rangle_u = v_P \left(
\begin{array}{ccc}
1 & 0 & 0 \\
0 & -1 & 0 \\
0 & 0 & 1
\end{array} \right).
\eqno(2.8)
$$
Here, in Eq.(2.8), the index ``$u$" denotes that 
a VEV matrix $\langle A\rangle$ takes a form 
$\langle A\rangle_u$ at the diagonal basis of 
the up-quark mass matrix $M_u$.
The reason of the existence of the matrix $\langle P_u \rangle_u$ 
is as follows: If we take a value $a_u \simeq -1.8$ in the up-quark
mass matrix relation given in Eq.(2.6), we can give reasonable 
up-quark mass ratios, but signs of the eigenvalues of 
$\langle \Phi_u \rangle$ show $(+,-,+)$,
i.e. $(M_u^{1/2})^{diag}\propto {\rm diag}(\sqrt{m_u},
-\sqrt{m_c}, \sqrt{m_t})$ for $a_u \simeq -1.8$.
In order to give reasonable neutrino mixing parameters 
(especially the observed value \cite{atm} 
$\sin^2 2\theta_{atm} \simeq 1$),
the existence of $P_u$ in Eq.(2.5) is indispensable 
in the prediction of $\sin^2 2\theta_{atm}$.
The $\xi_\nu$ term in Eq.(2.5) has been added from a reason that
the O(3) model was based on an O(3) family symmetry, so
that terms $ACB+BCA$ were also possible in addition to terms $ABC+CBA$.
By adjusting parameters $a_u$, $a_d$ and $\xi_\nu$,   
we have obtained \cite{O3_PLB09} not only the observed nearly 
tribimaximal neutrino mixing \cite{tribi}, but also reasonable 
Cabibbo-Kobayashi-Maskawa (CKM) quark mixing
(however, the fitting of CKM mixing is not so excellent
compared with that of neutrino mixing).

In conclusion, we summarize the previous O(3) model as follow:

\noindent
(i) The charged lepton mass matrix $M_e$ is given by a bilinear
form of a fundamental VEV matrix $\langle \Phi_e \rangle$, 
i.e. $(M_e)_{ij} \propto \langle (Y_e)_{ij} \rangle \propto
\langle (\Phi_e)_{ik} \rangle \langle (\Phi_e)_{kj} \rangle$.
This VEV matrix $\langle \Phi_e \rangle$ plays an important 
role in another mass matrices $M_u$, $M_d$ and $M_\nu$.

\noindent
(ii) All yukawaons $Y_f$ are singlets under the conventional 
gauge symmetries and they have the same O(3) assignments 
$({\bf 5}+{\bf 1})$, 
so that we assume an additional U(1) 
symmetry (we called it U(1)$_X$ symmetry), and each yukawaon
is distinguished from others by the U(1)$_X$ charge $Q_X$.

\noindent
(iii) In order to build a model without a Dirac neutrino 
yukawaon $Y_\nu$, we assume $Q_X(\nu^c)=Q_X(e^c)$, so that
the yukawaon $Y_e$ couple not only to the charged lepton 
sector, but also to the Dirac neutrino sector. 
As a result, the neutrino Majorana mass matrix $M_\nu$ is given
by a form $M_\nu = m_D M_R^{-1} m_D^T \propto 
\langle Y_e \rangle M_R^{-1} \langle Y_e^T \rangle$. 

\noindent
(iv)   We assume a form (2.5) as the form of $M_R$.
Even when we take only the dominant term 
$\Phi_u P_u Y_e + Y_e P_u \Phi_u$ into consideration, 
we can obtain 
favorable results $\sin^2 2\theta_{atm} \simeq 1$ 
and $|U_{13}|^2 \simeq 0$, although the predicted 
value $\tan^2 \theta_{solar} \simeq 0.7$ is somewhat
large compared with the observed value \cite{solar}
$\tan^2 \theta_{solar} \simeq 0.45$. 
(See  a case of $\xi_\nu=0$ in Table 1 in Sec.4.) 
Only when we take the $\xi_\nu$ term in Eq.(2.5) 
into consideration, we can fit the predicted 
$\tan^2 \theta_{solar}$ to the observed value 
without almost affecting the favorable results 
$\sin^2 2\theta_{atm} \simeq 1$ 
and $|U_{13}|^2 \simeq 0$.
We can also fit the neutrino mass ratios $R_\nu$ by 
adjusting the parameter $\xi_0$ in Eq.(2.5) 
without affecting the neutrino mixing parameters. 
 
\noindent (v)
The CKM parameter fitting in the original O(3) model
is not so excellent, although a rough tendency is fine.   
Somewhat improvements are needed. 
(For example, see Ref.\cite{N-K_PRD11}.)

\noindent (vi)
The model is based on an effective theory. 
The cutoff scale $\Lambda$ is of the order to $10^{12}$
GeV.  The scale is originated in the scale $M_R$ 
in the neutrino seesaw model.
As a result, it is hard to observe new physics which 
comes from the O(3) model. 

Correspondingly to the item numbers (i) - (vi) in the 
O(3) model, the present U(3)$\times$O(3) model have 
the following characteristics: 

\noindent 
(i) By considering $(M_e)_{ij} \propto 
\langle (Y_e)^{ij} \rangle \propto \langle (\Phi_e)^{i\alpha} \rangle 
\langle (\Phi_e^T)^{\alpha j} \rangle$, we can consider 
a similar picture described in (i) of the O(3) model.

\noindent
(ii) In the U(3)$\times$O(3) model, the yukawaons are 
assigned as $Y_e \sim ({\bf 6}^*, {\bf 1})$, 
$Y_R \sim ({\bf 1}, {\bf 5}+{\bf 1})$,
$Y_u \sim ({\bf 3}, {\bf 3})$ and 
$Y_d \sim ({\bf 1}, {\bf 5}+{\bf 1})$ of U(3)$\times$O(3).
Therefore, if we can assign $R$ charges to $Y_R$ and $Y_d$ 
differently, we do not need such U(1)$_X$ charges in the
O(3) model.

\noindent
(iii) In the O(3) model, in order to build a model 
without $Y_\nu$, we have assumed $Q_X(\nu^c)=Q_X(e^c)$
by hand.
In the present model, $Y_e^{i \alpha}$ couples to 
$\ell_i e^c_j$ as well as the O(3) model, while 
$(\Phi_e)^{i\alpha}$ couples only to $\ell_i \nu^c_\alpha$ 
without such a phenomenological assignment of the U(1) charge.  
As a result, the neutrino Majorana mass matrix $M_\nu$ is given
by a form $M_\nu \propto \langle \Phi_e \rangle M_R^{-1} 
\langle \Phi_e^T \rangle$. 

\noindent
(iv)   By replacing $Y_R$ in the form (2.5) with  
$\Phi_e Y_R \Phi_e^T$, we can obtain a practically 
same result as the dominant term $\Phi_u P_u Y_e + 
Y_e P_u \Phi_u$ in the form (2.5).
However, in the U(3)$\times$O(3) model, we cannot 
write such a term as $\xi_\nu$ in the O(3) model. 
Therefore, in the present model, we assume an alternative 
term which is invariant under the U(3)$\times$O(3)
symmetries. 
In spite of the existence of such a new $\xi_\nu$ term, 
as we show in Table 1 in Sec.4, 
 the new $\xi_\nu$ term can also give reasonable fits for
neutrino mixing parameters.
We can also fit the neutrino mass ratios $R_\nu$ by 
introducing a similar term to the $\xi_0$ term in Eq.(2.5). 
However, we do not discuss the numerical results of $R_\nu$ 
in this paper, because it is not prediction, but it is a result 
of adjusting the parameter $\xi_0$.

\noindent
(v) The numerical results for the CKM mixing parameters 
are identical with the 
results in the O(3) model, so that we do not show the results.
In order to obtain more precise agreements of the CKM mixing 
parameters with 
the observed values, further investigation is needed
for the present model, especially, for 
the structure of $Y_d$.

\noindent
(vi) The greatest merit of the present U(3)$\times$O(3) 
model is to lower the scale of $\Lambda$, e.g. 
$\Lambda \sim 10^8$ GeV, although $\Lambda \sim 10^{12}$
GeV in the O(3) model.
The details are discussed in Sec.5.

\vspace{2mm}

{\large\bf 3. Model for quark sector}

First, we investigate possible superpotential forms 
in the quark sector.
Correspondingly to the lepton sector with 
$(\ell_i, e^c_i, \nu_\alpha)$, we consider 
a model with $(q_\alpha,d^c_\alpha, u^{c i})$. 

We have introduced a field $\Phi^u_{i \alpha}$ similar to
$\Phi_e^{i\alpha}$ as shown in Eq.(1.4).
However, we cannot identify $\Phi^u_{i\alpha}$ as $Y^d_{\alpha\beta}$ 
although we have regarded $\Phi_e$ as $Y_\nu^{Dirac}$ in the
lepton sector. 
Note that in this model, the U(3) gauge bosons $A_i^j$ 
which couple to charged lepton sector cannot couple to 
the down-quark sector. 
For yukawaons in the quark sector, we assume the following superpotential 
terms:\footnote{
In order to distinguish $\Phi^u_{i\alpha}$ from 
$Y^u_{i\alpha}$, $Y^d_{\alpha\beta}$ from 
$Y^R_{\alpha\beta}$, and so on, we must assume 
different $R$ charges for those fields.
For the $R$-charge assignments, see Table 2 in Sec.6.
} 
$$
W_u = \left( \mu_u Y^u_{i\alpha} 
+ \frac{\lambda_u}{\Lambda}
\Phi^u_{i\beta} \bar{E}_{u3}^{T\beta j} \Phi^{u}_{ j\alpha} \right)
 \Theta_u^{\alpha i} 
+  \frac{1}{\Lambda} \left( \lambda'_u \bar{E}_{u6}^{ik}
\Phi^u_{k\alpha} \bar{E}_{u3}^{T\alpha j} 
+ {\lambda^{\prime\prime}_u}
\Phi_e^{i \alpha}(E_{\alpha\beta}+a_u S_{ \alpha \beta} ) 
\Phi_e^{T \beta j} \right) \Theta^{u \prime}_{ji}  ,
\eqno(3.1)
$$
$$
W_d = \frac{1}{\Lambda}  \left( \lambda_d
\bar{P}_d^{i\alpha} Y^d_{\alpha\beta} \bar{P}_d^{T\beta j} 
 + \lambda'_d \Phi_e^{i\alpha} (E_{\alpha\beta}+a_u S_{ \alpha \beta} )
\Phi_e^{T \beta j} \right) \Theta^d_{ji} ,
\eqno(3.2)
$$
where the VEV forms of $\langle \bar{E}_{u3}\rangle$, 
$\langle \bar{E}_{u6}\rangle$, $\langle E\rangle$ and 
$\langle S \rangle$ are given by $\langle \bar{E}_{u3}\rangle = 
v_{Eu3} {\bf 1}$, $\langle \bar{E}_{u6}\rangle = v_{Eu6} {\bf 1}$, 
$\langle E\rangle = v_E {\bf 1}$
and $\langle S \rangle = v_S X$ [$X$ is defined by Eq.(2.7)]. 
These forms of the superpotential effectively lead to the 
quark mass matrices given in Eq.(2.6) in the O(3) model.

The form $\langle S \rangle = v_S X$ is given by the following 
superpotential term $W_S$:
$$
W_S =\lambda_S \, {\rm det}S .
\eqno(3.3)
$$
By using a formula for any $3\times 3$ Hermitian matrix $A$
$$
{\rm det}A = \frac{1}{3} {\rm Tr}[AAA] -\frac{1}{2} {\rm Tr}[AA] {\rm Tr}[A]
+\frac{1}{6} ({\rm Tr}[A])^3 ,
\eqno(3.4)
$$
we obtain
$$
\frac{\partial W_S}{\partial S} = \lambda_S \left[ SS 
-\frac{1}{2} \left( 2 S  {\rm Tr}[S] + {\bf 1}  {\rm Tr}[SS] \right)
+\frac{1}{2} {\bf 1} ( {\rm Tr}[S])^2 \right] .
\eqno(3.5)
$$
Therefore, the SUSY vacuum condition $\partial W_S/\partial S =0$ leads
to a solution
$$
\langle S \rangle \langle S \rangle = \langle S \rangle{\rm Tr}[ \langle S \rangle] .
\eqno(3.6)
$$
By applying another formula for any $3\times 3$ Hermitian matrix $A$
$$
{\bf 1} {\rm det}A = AAA - AA {\rm Tr}[A] + \frac{1}{2} A \left[
({\rm Tr}[A])^2 - {\rm Tr}[AA] \right] ,
\eqno(3.7)
$$
to the solution (3.6), we obtain 
$$
{\rm det}\langle S \rangle=0 .
\eqno(3.8)
$$
Therefore, from Eqs.(3.6) and (3.8), we choose a specific form
$$
\langle S \rangle_e  = v_S X = \frac{1}{3} v_S \left(
\begin{array}{ccc}
1 & 1 & 1 \\
1 & 1 & 1 \\
1 & 1 & 1
\end{array}
\right) .
\eqno(3.9)
$$
[Of course, the solution (3.9) is not a general solution.
For example, $\langle S \rangle  = v_S \, {\rm diag}(0,0,1)$ is
also possible.
We assume that $\langle S \rangle_e$ is given by the form
(3.9) in the diagonal basis of the charged lepton 
mass matrix $M_e$ (i.e. of $\langle \Phi_e\rangle_e$).]
On the other hand, for the form $\langle E\rangle_e = v_E \, {\bf 1}$,
we consider a superpotential form
$$
W_E = \lambda_E {\rm Tr}[EE] {\rm Tr}[E] 
+\lambda'_E({\rm Tr}[E] )^3 .
\eqno(3.10)
$$
A SUSY vacuum condition 
$$
\frac{\partial W_E}{\partial E} = 2\lambda_E E \, {\rm Tr}[E] +
{\bf 1} \left\{ \lambda_E {\rm Tr}[EE] +3 \lambda'_E ({\rm Tr}[E])^2
\right\} = 0,
\eqno(3.11)
$$
leads to $\langle E\rangle = v_E \, {\bf 1}$.
Since we require the $R$-charge conservation, we have to assign 
$R=2/3$ to the fields $S$ and $E$.

In general, for fields $A_{i\alpha}$ and $\bar{A}^{\alpha i}$
with $R(A) +R(\bar{A}) =1$, we can consider superpotential terms
$$
W_A =\frac{1}{\Lambda} \left( \lambda_A {\rm Tr}[A\bar{A} A\bar{A}]
+ \lambda'_A {\rm Tr}[A\bar{A}] {\rm Tr}[A\bar{A}] \right) .
\eqno(3.12)
$$
SUSY vacuum conditions $\partial W_A/\partial A=0$ and
$\partial W_A/\partial \bar{A}=0$ lead to
$$
A \bar{A} = - \frac{\lambda'_A}{\lambda_A} {\bf 1}
 {\rm Tr}[A\bar{A}] .
 \eqno(3.13)
$$
We assume that $\bar{E}_{u3}^{i\alpha}$, $\bar{E}_{u6}^{ij}$ and 
$\bar{P}_d^{i\alpha}$ are given by a 
superpotential form (3.12).
We take a special solution 
$$
\langle A \rangle = a \,{\rm diag}(e^{i\phi_1},  e^{i\phi_2}, e^{i\phi_3}),
\eqno(3.14)
$$
in $\langle \bar{A} \rangle \langle A \rangle \propto {\bf 1}$, 
i.e. $\langle E^{u3}\rangle = v_{Eu3} {\bf 1}$, 
$\langle E^{u6}\rangle = v_{Eu6} {\bf 1}$ and 
$\langle \bar{P}_d \rangle = v_{Pd} \,{\rm diag}(e^{i\phi_1},  
e^{i\phi_2}, e^{i\phi_3})$.
(In any case, we must assume a special form of 
$\langle \bar{A} \rangle$ in $\langle \bar{A} \rangle \langle A \rangle
\propto {\bf 1}$.)

In the O(3) model, the CKM parameter fitting is not so remarkable 
compared with those in the up-quark mass ratios and the neutrino 
mixing parameters.
Note that, differently from the O(3) model, we have the phase matrix 
$\langle \bar{P}_d \rangle$ as seen in Eq.(3.2).
This will improve the fitting for the CKM mixing parameters. 
(However, such a numerical fitting is not our purpose 
in the present paper.)

\vspace{2mm}

{\large\bf 4. Neutrino mass matrix}

In the present model, since $m_D \propto  \langle \Phi_e \rangle$ 
differently from the O(3) model, we cannot consider the form 
$Y_R = \Phi_u P_u Y_e +Y_e P_u \Phi_u$ as shown in Eq.(2.5).
Therefore, we consider the following superpotential terms for 
the $Y_R$ sector:
$$
\frac{1}{\Lambda} \left[ \lambda_R 
\Phi_e^{i \alpha} Y_R^{\alpha \beta} \Phi_e^{T\beta j}
+\lambda^{\prime}_R \left( 
\bar{P}_u^{i\alpha} \Phi^{Tu}_{\alpha k} Y_e^{kj} +
Y_e^{ik} \Phi^{u}_{k \alpha } \bar{P}_u^{T \alpha j}  \right) 
\right] \Theta^R_{ji} .
\eqno(4.1)
$$ 
Note  that, in the present model, we cannot consider such a 
term which corresponds to the $\xi_\nu$ term in the O(3) model  
[Eq.(2.5)].
In the O(3) model, in order to give the observed value \cite{solar}
$\tan^2 \theta_{solar} \simeq 1/2$, it was indispensable 
that we take a non-vanishing value of $\xi_\nu$, 
although  we could give the observed values \cite{atm} 
 $\sin^2 2\theta_{atm}\simeq 1$ and $|U_{13}|^2 \simeq 0$ 
even if $\xi_\nu=0$.  
Therefore, in the present model, too, we need some additional 
term to Eq.(4.1). 
We assume the following 
superpotential for $Y_R$ with a new $\xi_\nu$ term:
$$
W_R =\frac{1}{\Lambda} \left\{
\lambda_R (\Phi_e Y_R \Phi_e^T)^{ij} 
+\lambda'_R \left[ (\bar{P}_u \Phi_u^T)^i_k Y_e^{kj}
+Y_e^{ik}(\Phi_u \bar{P}_u^T)^j_k  \right. \right.
$$
$$
\left.\left. +
\xi_\nu (\Phi_u \bar{P}_u^T)^k_k Y_e^{ij} \right] 
+ \lambda_R^{\prime\prime} Y_e^{ik} E_{kl}^{u6} Y_e^{lj} 
\right\} \Theta_{ji}^R .
\eqno(4.2)
$$
The last term ($ \lambda_R^{\prime\prime}$ term) has been added 
in order to adjust the neutrino mass ration $R_\nu =\Delta m^2_{solar}
/\Delta m^2_{atm}$ similar to the $\xi_0$ term in Eq.(2.5) in the 
O(3) model.
From the superpotential (4.2), we obtain the following neutrino 
mass matrix $M_\nu$:
$$
M_\nu \propto  \langle \Phi_e \rangle \left\{ 
\langle \Phi_e \rangle^{-1} \left[\langle P_u \rangle 
\langle \Phi_u \rangle \langle Y_e \rangle 
+\langle Y_e \rangle \langle \Phi_u \rangle \langle P_u \rangle 
+ \xi_\nu {\rm Tr}[\langle \Phi_u \rangle \langle P_u \rangle 
\right. \right.
$$
$$
\left. \left.
+ \xi_0 \langle Y_e\rangle \langle E^{u6}\rangle \langle Y_e\rangle ] 
\langle Y_e \rangle \right] \langle \Phi_e \rangle^{-1}
\right\}^{-1} \langle \Phi_e \rangle ,
\eqno(4.3)
$$
where $\langle \Phi_e\rangle_e \propto 
{\rm diag}(\sqrt{m_e}, \sqrt{m_\mu}, \sqrt{m_\tau})$
and $\langle \Phi_u\rangle_e = k_u  
\langle \Phi_e\rangle_e (\langle E\rangle + a_u 
\langle S_u\rangle_e )\langle \Phi_e\rangle_e $.

\begin{table}
\begin{center}
\begin{tabular}{cccc} \hline
$\xi_\nu$ & $\tan^2 \theta_{solar}$ & $\sin^2 2\theta_{atm}$
& $|U_{13}|^2$ \\ \hline
$0$    & $0.6995$  & $0.9872$ & $1.72\times 10^{-4}$ \\
$0.009$ & $0.4610$ & $0.9902$ & $2.28\times 10^{-4}$ \\
$0.010$ & $0.4408$ & $0.9905$ & $2.35\times 10^{-4}$ \\
\hline
\end{tabular}  
\end{center}
\begin{quotation}
\caption{
$\xi_\nu$ dependence of the neutrino parameters.
The value of $a_u$ is taken as $a_u=-1.78$
which can give reasonable up-quark mass ratios.
}
\end{quotation}
\end{table}

In Table 1, we demonstrate $\xi_\nu$-dependence
of the neutrino mixing parameters in a case with
$a_u=-1.78$, which gives reasonable up-quark mass
ratios 
$$
\sqrt{\frac{ {m_u}}{{m_c}}} = 0.04389, \ \ \ \   
\sqrt{\frac{ {m_c}}{m_t}} = 0.05564.
\eqno(4.4)
$$
The predicted values are in good agreement with 
the observed values at $\mu=m_Z$ \cite{q-mass} 
$\sqrt{ {m_u}/{m_c}} = 0.045^{+0.013}_{-0.010}$ and  
$\sqrt{ {m_c}/{m_t}} = 0.060 \pm 0.005$.
As seen in Table 1, the magnitudes of 
$\sin^2 2\theta_{atm}$ and $|U_{13}|^2$ are almost
independent of the parameter $\xi_\nu$ (e.g.  
$\sin^2 2\theta_{atm} \simeq 1$ and $|U_{13}|^2
\simeq O(10^{-4}$), while
the value of $\tan^2 \theta_{solar}$ is highly 
dependent on the value of $\xi_\nu$ (e.g. 
$\tan^2 \theta_{solar}=0.70$ for $\xi_\nu =0$ and 
$\tan^2 \theta_{solar}=0.44$ for $\xi_\nu =0.01$). 
We note that the model can give
excellent fits with the observed values of the
neutrino mixing parameters in spite of
its phenomenologically different form of the 
$\xi_\nu$ term.

\vspace{2mm}

{\large\bf 5.  Energy scale of the cutoff $\Lambda$}

So far, we did not discuss the energy scale $\Lambda$
of the effective theory.
In this section, we discuss that we can lower 
the value of $\Lambda$ in the present model. 

In the O(3) model (also in the present model, too), the charged 
lepton mass matrix $M_e$ is given by
$$
(M_e)_{33} = y_e \frac{\langle (Y_e)_{33} \rangle}{\Lambda} 
\langle H_d^0 \rangle,
\eqno(5.1)
$$ 
so that, in order to give $m_\tau \sim 1$ GeV with 
$\langle H_d^0 \rangle \sim 10$ GeV ($\tan\beta \sim 10$),
we have to take ${\langle Y_e\rangle}/{\Lambda} \sim 10^{-1}$.
Since the neutrino mass matrix in the O(3) model
is given by
$$
(M_\nu)_{33} =(m_D)_{3k} (M_R^{-1})_{kl} (m_D)_{l3} 
= \frac{(y_\nu)^2}{\lambda_R} 
\langle H_u^0 \rangle^2 \frac{1}{\Lambda^2}
\langle (Y_e)_{33} \rangle (\langle Y_R\rangle^{-1})_{33} 
(\langle Y_e\rangle)_{33} ,
\eqno(5.2)
$$
we have to take $\langle Y_R\rangle \sim 10^{12}$ GeV in order to
give $m_{\nu 3}/m_\tau \sim 10^{-10}$.

On the other hand, in contrast to the O(3) model, 
the neutrino mass matrix in the present model is given by
$$
(M_\nu)_{33} =(m_D M_R^{-1} m_D^T)_{33} = \frac{(y_\nu)^2}{\lambda_R} 
 \frac{1}{\Lambda^2} 
\langle H_u^0 \rangle^2 \langle (\Phi_e)^{33} \rangle 
(\langle Y_R\rangle^{-1})_{33} \langle (\Phi_e)^{33} \rangle ,
\eqno(5.3)
$$
and the charged lepton mass matrix is given by
$$
(M_e)_{33} = y_e \frac{\langle (Y_e)^{33} \rangle}{\Lambda} 
\langle H_d^0 \rangle
= - y_e \lambda_e \frac{\Lambda}{\mu_e} 
\frac{\langle (\Phi_e)^{33} \rangle}{\Lambda}
\frac{\langle (\Phi_e)^{33} \rangle}{\Lambda}\langle H_d^0 \rangle ,
\eqno(5.4)
$$
from Eq.(2.3).
Therefore, we can estimate 
$$
\frac{(M_{\nu })_{33} }{m_\tau} 
= \frac{(y_\nu)^2}{y_e \lambda_e \lambda_R}  
\frac{ \langle H_u^0\rangle^2}{\langle H_d^0 \rangle} 
\left(  \langle Y_R^{-1}\rangle \right)_{33} 
\frac{\mu_e}{\Lambda}.
\eqno(5.5)
$$
By taking $(M_{\nu })_{33}/m_\tau \sim 10^{-10}$ 
(by regarding $(M_{\nu })_{33}$ as $(M_{\nu })_{33}
\sim m_{\nu 3} \simeq \sqrt{\Delta m_{atm}^2} \simeq 0.047$ eV 
\cite{atm}), 
$\langle H_u^0\rangle \sim 10^2$ GeV and 
$\langle H_d^0\rangle \sim 10$ GeV, and by assuming
$\langle Y_R\rangle \sim \Lambda$ (a maximum value of
$\langle Y_R \rangle$ in the present effective theory) , 
we obtain
$$
\Lambda \sim \sqrt{ 10^{13}[{\rm GeV}]\, \mu_e[{\rm GeV}]}.
\eqno(5.6)
$$
Therefore, we can, in principle, take any small 
value of $\Lambda$ by assuming a small value of $\mu_e$.
(However, a too small value of $\mu_e$ is unlikely.)
For example, if we take $\mu_e \sim 1$ TeV, 
we obtain $\Lambda \sim 10^8$ GeV. 

On the other hand, the U(3) gauge bosons $A_i^j$ 
acquire their masses from VEVs $\langle Y_e^{ij}\rangle$,
$\langle \Phi_e^{i\alpha}\rangle$,  
$\langle Y^u_{i\alpha}\rangle$, and so on.
In general, the gauge boson masses $m(A_i^j)$ are given by
$$
m^2(A_i^j) = \frac{1}{2} \left[ \sum_{\bf 6} |\langle {\bf 6}_i
\rangle +\langle {\bf 6}_j \rangle|^2 + \sum_{\bf 3}
\left( |\langle {\bf 3}_i\rangle|^2 +|\langle {\bf 3}_j \rangle|^2
\right) \right] ,
\eqno(5.7)
$$
where $\langle {\bf 6}_i\rangle$ and $\langle {\bf 3}_i\rangle$ are 
eigenvalues of VEV matrices of fields ${\bf 6}$ (${\bf 6}^*$) and 
${\bf 3}$ (${\bf 3}^*$), respectively, and the observed CKM 
mixing has been neglected.
At present, magnitudes of $\langle \bar{E}_{u6}\rangle$, 
$\langle \bar{E}_{u3}\rangle$, and so on, are free parameters. 
If we consider that the dominant contributions to $m(A_i^j)$ are, 
for example, $\langle \bar{E}_{u6}^{ij} \rangle = \delta_{ij} \Lambda$,
we obtain $m(A_i^j) \simeq \sqrt{2} g \Lambda$, so that we cannot 
observe the gauge boson effects.
Here, let us optimistically suppose that dominant 
contributions are $\langle Y_u^{ii} \rangle \sim (m_{ui}/
\langle H_u^0 \rangle) \Lambda$.
(We can easily demonstrate that $\langle Y_u \rangle$ is dominant,
at least, compared with $\langle Y_e \rangle$ and 
$\langle \Phi_e \rangle$.)
Then, we can estimate the gauge boson masses as follows:
$$
m(A_i^j) \simeq \frac{g}{\sqrt2} \sqrt{(v_i^u)^2+(v_j^u)^2}
= \frac{g}{\sqrt2} \sqrt{m_{ui}^2+m_{uj}^2}\frac{ \Lambda}{ y_u 
\langle H_u^0\rangle }  ,
\eqno(5.8)
$$
where we have taken a flavor basis which is defined by 
$\langle Y^u_{i\alpha}\rangle = \delta_{i\alpha} v^u_i$,
so that we can estimate gauge boson mass ratios as  
$$
\frac{m(A_1^1)}{m(A_3^3)} \simeq \frac{m_u}{m_t} 
\simeq 0.74 \times 10^{-5} , \ \ \ \ 
\frac{m(A_2^1)}{m(A_3^3)} \simeq \frac{m_c}{\sqrt2 m_t} 
\simeq 2.6 \times 10^{-3} ,
\eqno(5.9)
$$
where $m(A_3^3) \sim \Lambda$.
If we suppose $\Lambda \sim 10^8$ GeV, we obtain
$m(A_1^1) \sim 1$ TeV and $m(A_2^1) \sim 10^2$ TeV.
The gauge boson $A_1^1$ with $m(A_1^1) \sim 1$ TeV  will be 
observed in $Z'$ searches at the LHC at which $Z'$ can 
decay into $e^+ + e^-$ but not into $\mu^+ + \mu^-$.
(More details of the $A_1^1$ search at LHC have substantially 
been discussed in Ref.\cite{KSY_PLB11}.)
Since the U(3) gauge bosons cannot couple to the 
down-quark sector and do only to $u^c$, the gauge 
boson $A_2^1$ with the next lower mass can  
contribute to the $D^0$-$\bar{D}^0$ mass difference.
In this paper, we do not give further numerical predictions.
Phenomenological meanings of the present model
in TeV region physics will be discussed elsewhere. 
We again would like to emphasize that such a lower 
scale of $\Lambda$ would not be realized without 
introducing the yukawaon $Y_e^{ij}$ which is 
related to $\Phi_e^{i\alpha}\Phi_e^{T \alpha j}$ 
by Eq.(2.3).

\vspace{2mm}

{\large\bf 6. Concluding remarks}

In conclusion, stimulated by the Sumino's model 
\cite{Sumino09PLB,Sumino09JHEP}
with U(3)$\times$O(3) symmetries for the charged
leptons, we have considered 
a SUSY version of his model with a U(3) family gauge 
symmetry by extending it to U(3)$\times$O(3) family
symmetries for  the quarks and leptons 
(but O(3) is already broken at $\mu=\Lambda$).
Although, in order to distinguish each yukawaon 
from other ones, an additional U(1) 
symmetry [i.e. U(1)$_X$] was assumed in the previous O(3) 
yukawaon model, we do not need such U(1)$_X$ charges 
in the present model with the two family symmetries 
(but we still assume the $R$ charge conservation).  
Quantum number assignments of the fields are summarized 
in Table 2.  
Since we consider a superpotential term $\mu_H H_u H_d$,
$R$ charges for Higgs fields $H_u$ and $H_d$ satisfy 
$R(H_u)+R(H_d)=2$.
For simplicity, in Table 2, we have taken  $R(H_u)=R(H_d)=1$.
Furthermore, we have taken $R(\ell)=R(q)=1$. 
In spite of such simplified assignments, we still have 
free parameters in the $R$ charge assignments as seen
in Table 2.
In an explicit $R$ charge assignments, we must take care 
that yukawaon fields with the same U(3)$\times$O(3) assignments 
cannot have the same $R$ charges, because those are  
singlets under the conventional gauge symmetry, and they should 
distinguished from  each other only by $R$ charges.

\begin{table}
\begin{center}
\begin{tabular}{|c|cccccccccc|} \hline
   & $\ell$ & $e^c$ & $\nu^c$ & $q$  & $u^c$ & $d^c$ &
$H_u$ & $H_d$ & $Y_e$ & $\bar{E}_{u6}$  \\ \hline
U(3) & ${\bf 3}$ & ${\bf 3}$ & ${\bf 1}$ & ${\bf 1}$ &
 ${\bf 3}^*$ & ${\bf 1}$ & ${\bf 1}$ & ${\bf 1}$ & 
${\bf 6}^*$ & ${\bf 6}^*$  \\
O(3) & ${\bf 1}$ & ${\bf 1}$ & ${\bf 3}$ & ${\bf 3}$ &
 ${\bf 1}$ & ${\bf 3}$ & ${\bf 1}$ & ${\bf 1}$ & 
${\bf 1}$ & ${\bf 1}$   \\
 $R$ & $1$ & $-2r_e$ & $-r_e$ & $1$ & $-(2 r_u+\bar{r}_{u3})$ & 
$-r_d$ & 
$1$ & $1$ &  $2 r_e$ & $ 1$ \\ \hline
\end{tabular} 

\begin{tabular}{|cccccccccc|} \hline
$\Theta^e$ & $\Theta^R$  & $\Theta^{u\prime}$ & $\Theta^d$ & $E^{u6}$ & 
$\Phi_e$ & $\bar{P}_u$ & $\bar{P}_d$ & $\Theta_u$ & $\bar{E}_{u3}$  
\\ \hline
${\bf 6}$ & ${\bf 6}$ & ${\bf 6}$ & ${\bf 6}$ & ${\bf 6}$ &
${\bf 3}^*$ & ${\bf 3}^*$ & ${\bf 3}^*$ & ${\bf 3}^*$ & ${\bf 3}^*$ \\
${\bf 1}$ & ${\bf 1}$ & ${\bf 1}$ & ${\bf 1}$ & ${\bf 1}$ &
${\bf 3}$ & ${\bf 3}$ & ${\bf 3}$ & ${\bf 3}$ & ${\bf 3}$ \\
$2-2 r_e$ & $R(\Theta^R)$ & $R(\Theta^{u\prime})$ & $R(\Theta^d)$ & 
$0$ & $r_e$ & $\bar{r}_{Pu}$  & $\bar{r}_{Pd}$  & 
$R(\Theta_u)$ & $\bar{r}_{u3}$ \\ \hline
\end{tabular}

\begin{tabular}{|ccccccc|} \hline
$Y^u$ & $\Phi^u$ & $P^d$ &  $E^{u3}$ & 
$Y_R$ & $Y_d$ & $S$ \& $E$ \\ \hline
${\bf 3}$ & ${\bf 3}$ & ${\bf 3}$ & ${\bf 3}$ & 
${\bf 1}$ & ${\bf 1}$ & ${\bf 1}$ \\
${\bf 3}$ & ${\bf 3}$ & ${\bf 3}$ & ${\bf 3}$ &
${\bf 5}+{\bf 1}$ & ${\bf 5}+{\bf 1}$ & ${\bf 5}+{\bf 1}$ \\
$2 r_u + \bar{r}_{u3} $ & $r_u$ & $1-\bar{r}_{Pd}$ & $1-\bar{r}_{u3}$ &
$2r_e$ & $r_d$ & $\frac{2}{3}$ \\ \hline
\end{tabular}

\end{center}
\begin{quotation}
\caption{
Quantum numbers of U(3)$\times$O(3) family symmetries and $R$ charges, 
where $R(\Theta^R)=2 -4r_e$, $R(\Theta^{u\prime})=1-(r_u 
+ \bar{r}_{u3})$, $R(\Theta^d)=2-(r_d +2 \bar{r}_{Pd})$ 
and $R(\Theta_u)=2 -(2 r_u +\bar{r}_{u3})$. 
For simplicity, we have taken as $R(H_u)=R(H_d)=1$ and 
$R(\ell)=R(q)=1$. 
Here, relations $\bar{r}_{Pu}=2 r_e-r_u$ and 
$2 r_e+\frac{2}{3} = r_u + \bar{r}_{E3} +1 =
r_d + 2 \bar{r}_{Pd}$ are required. 
The value of $R(E^{u6})$ is fixed as $R(E^{u6})=0$ due to the 
relation $R(Y_R)+2 R(\Phi_e)=2R(Y_e)+R(E^{u6})$ term in Eq.(4.2). 
}
\end{quotation}
\end{table}


The original Sumino model is a model for the 
charged leptons, so that he has explicitly mentioned 
nothing as to quark and neutrino sectors.
If we adopt Sumino's assignments $\ell \sim {\bf 3}$
and $e^c \sim {\bf 3}$ of U(3), while we assume 
$\nu^c \sim {\bf 3}$ of O(3), 
the model leads to a seesaw-type 
neutrino mass matrix 
$M_\nu =\langle \Phi_e \rangle M_R^{-1} \langle \Phi_e \rangle^T$. 
We have investigated a possible form of the
Majorana mass matrix $M_R= \lambda_R \langle
Y_R\rangle$ of the right-handed neutrinos 
by referring to a supersymmetric yukawaon
model (O(3) model) \cite{O3_PLB09} for the 
neutrino sector.
The present form (4.2) of $M_R$ is similar to the
form  (2.5) in the O(3) model, but the $\xi_\nu$ 
term is completely different from the O(3) model.
Nevertheless, in this model, too, we can
successfully obtain the nearly 
tribimaximal neutrino mixing by adjusting 
the parameter $\xi_\nu$ as seen in Table 1.
It is worthwhile noticing this.

As seen in Table 2, we have two ${\bf 6}^*$, five ${\bf 6}$,
five ${\bf 3}^*$ and four ${\bf 3}$ of U(3) except for 
the quarks and leptons.
Therefore, the present model is not anomaly free. 
At present, the present model for $M_d$ cannot give 
precise numerical fits for the observed CKM mixing.
The improvement of $Y_d$ structure is an open question 
at present.
For such improvement, we will need further fields which
are singlets under SU(3)$_c \times$SU(2)$_L\times$U(1)$_Y$, 
but non-singlets under family symmetries. 
Inversely, since we have too many yukawaons, some of them
will be economized in future.
Although we consider that the theory should be anomaly 
free, at this stage of the yukawaon model, it will not 
be fruitful to adhere to the anomaly freedom problem.

The greatest merit in considering U(3)$\times$O(3) 
family gauge symmetries lies in that we can lower
the cutoff scale $\Lambda$ in the present yukawaon model.
In the previous yukawaon model, in order to give the 
observed tiny neutrino masses, we had been obliged 
to consider $\Lambda \sim 10^{12}$ GeV.
In the O(3) model, the relation (2.4) was ad hoc assumed
(i.e. the yukawaon $Y_\nu$ was regarded as $Y_e$).
In the present model, the field $\Phi_e$ can couple
to the neutrino Dirac term because of the same quantum
numbers of U(3)$\times$O(3), so that we obtain Eq.(5.3)
instead of the relation (5.2). 
This has enabled us to lower the scale $\Lambda$ 
as we have seen in Sec.5. 

If we suppose a value $\mu_e \sim 1$ TeV which gives 
$\Lambda \sim 10^8$ GeV, we obtain $m(A_1^1) \sim 1$ TeV 
and $m(A_2^1) \sim 10^2$ TeV in an optimistic case.
The gauge boson $A_1^1$ with $m(A_1^1) \sim 1$ TeV  will be 
observed in $Z'$ searches at the LHC at which $Z'$ can 
decay into $e^+ + e^-$ but not into $\mu^+ + \mu^-$.
The gauge boson $A_2^1$ with the next lower mass
can contribute to $D^0$-$\bar{D}^0$ mass difference.
Phenomenological studies of the present model
in TeV region physics will be discussed elsewhere. 
We would like to emphasize that, in order to make 
yukawaon effects visible in the terrestrial experiments, 
it has been inevitable to adopt the present model with 
two family symmetries.  


\vspace{10mm}
{\Large\bf Acknowledgments}   

The authors would like to thank Y.~Sumino and
M.~Yamanaka for enjoyable conversations. 
He also thank T.~Yamashita for helpful comments
on anomaly-free conditions and effective 
theories, and especially for pointing out an mistake
in an earlier version of this paper, and J.~Kubo for 
useful comments for a FCNC problem. 
The work is supported by JSPS 
(No.\ 21540266).



\end{document}